\documentclass[12pt]{article}

\usepackage{amsmath,amssymb,amsfonts,graphics,graphicx,amscd,amsfonts,epsfig,color}
\usepackage{subfig}
\usepackage{cite}
\usepackage{enumitem}
\usepackage{here}

\newcommand {\beq} {\begin{equation}}
\newcommand {\eeq} {\end{equation}}
\newcommand {\beqa}{\begin{eqnarray}}
\newcommand {\eeqa}{\end{eqnarray}}
\newcommand {\nn} {\nonumber}
\newcommand {\del} {\partial}

\newcommand {\Tr}{\mbox{Tr\,}}

\newcommand {\ee}{\mbox{e}}

\makeatletter
    
    \@addtoreset{equation}{section}
  \makeatother

\allowdisplaybreaks[4]

\date{}
\begin{document}

\begin{flushright} 
KEK-TH-2119
\end{flushright} 

\vspace{0.1cm}

\begin{center}
{\LARGE Complex Langevin analysis of the space-time structure} \\[1mm]
{\LARGE  in the Lorentzian type IIB matrix model}
\end{center}

\vspace{0.1cm}
\vspace{0.1cm}
\begin{center}

         Jun N{\sc ishimura}$^{ab}$\footnote
          {
 E-mail address : jnishi@post.kek.jp} 
and
         Asato T{\sc suchiya}$^{c}$\footnote
          {
 E-mail address : tsuchiya.asato@shizuoka.ac.jp} 

\vspace{0.5cm}

$^a${\it KEK Theory Center, 
High Energy Accelerator Research Organization,\\
1-1 Oho, Tsukuba, Ibaraki 305-0801, Japan} 

$^b${\it Department of Particle and Nuclear Physics, 
School of High Energy Accelerator Science,\\
Graduate University for Advanced Studies (SOKENDAI),\\
1-1 Oho, Tsukuba, Ibaraki 305-0801, Japan} 

$^c${\it Department of Physics, Shizuoka University,\\
836 Ohya, Suruga-ku, Shizuoka 422-8529, Japan}

\end{center}

\vspace{1.5cm}

\begin{center}
  {\bf abstract}
\end{center}

\noindent 
The Lorentzian type IIB matrix model has been 
studied as a promising candidate for 
a nonperturbative formulation of superstring theory.
In particular, the emergence of (3+1)D expanding space-time
was observed by Monte Carlo studies of this model.
It has been found recently, however, that the matrix configurations
generated by the simulation is singular
in that the submatrices representing the expanding 3D space
have only two large eigenvalues associated with the Pauli matrices.
This problem has been attributed to the approximation used
to avoid the sign problem in simulating the model.
%
Here we investigate the model
using the complex Langevin method to overcome the sign problem
instead of using the approximation.
Our results indicate
a clear departure from the Pauli-matrix structure,
while the (3+1)D expanding behavior is 
kept intact.

\newpage

\section{Introduction}


Nonperturbative studies often provide totally new perspectives
in quantum field theories.
Confinement of quarks, for instance, has been 
vividly demonstrated by the strong coupling expansion 
in the lattice gauge theory \cite{Wilson:1974sk}.
We consider the emergence of (3+1)D expanding space-time
in the Lorentzian version of the type IIB matrix model is 
another case of this sort \cite{Kim:2011cr}.
This model was conjectured to be a nonperturbative formulation of
superstring theory \cite{Ishibashi:1996xs} 
analogous to the lattice gauge theory in QCD.
The model has ten bosonic $N\times N$ Hermitian matrices, which represent
ten-dimensional space-time in the large-$N$ limit.
One of the most interesting features is that
the eigenvalue distribution of the ten bosonic
matrices can collapse to a lower-dimensional manifold, which
may be interpreted as 
the actual space-time dynamically generated in this model.
If this really happens,
it implies that
the (9+1)D
Lorentz symmetry of the model is spontaneously broken.

Monte Carlo studies of the type IIB matrix model
are extremely hard, however,
due to the so-called sign problem
caused by the complex weight in the partition function.
In the Eulicidean version, it comes from the Pfaffian
that is obtained by integrating out fermionic matrices,
while in the Lorentzian version, it comes from
the phase factor $\ee^{iS_{\rm b}}$ with the bosonic action $S_{\rm b}$.
If we treat the phase of the complex weight by reweighting,
huge cancellation among configurations with different phases occurs,
which makes the calculation impractical.
Recently the complex Langevin 
method (CLM) \cite{Parisi:1984cs,Klauder:1983sp}
has been attracting much attention
as a promising approach to this 
problem \cite{Aarts:2009dg,Aarts:2009uq,Aarts:2011ax,Nishimura:2015pba,Nagata:2015uga,Nagata:2016vkn,Ito:2016efb}.
In particular, it has been applied
successfully to the Euclidean version of 
the 6D type IIB matrix model \cite{Anagnostopoulos:2017gos},
and the spontaneous breaking of
the rotational SO(6) symmetry to SO(3) suggested 
by the Gaussian expansion method \cite{Aoyama:2010ry} has been confirmed.

In ref.~\cite{Kim:2011cr} and our subsequent
work \cite{Ito:2013ywa,Ito:2015mxa,Ito:2017rcr,Azuma:2017dcb}
on the Lorentzian type IIB matrix model and its simplified models,
the sign problem was avoided by
integrating out the scale factor
of the bosonic matrices by hand, which yields a function of
the bosonic action $S_{\rm b}$ sharply peaked at the origin.
Approximating this function by a sharply peaked Gaussian function,
we can perform Monte Carlo simulations without the sign problem.
The emergence of (3+1)D expanding space-time
was obtained in this way \cite{Kim:2011cr}.
The expanding behavior for a longer time was 
investigated by simulating the simplified models.
The obtained results suggested a scenario for the full model 
that the expansion is exponential
at early times \cite{Ito:2013ywa},
which is reminiscent of the inflation,
and that it turns into a power law \cite{Ito:2015mxa}
at later times, which is reminiscent of
the Friedmann-Robertson-Walker universe in the radiation dominated era.
See also refs.~\cite{Yang:2015vna,Kim:2018mfv,Tomita:2015let}
for closely related work.

It has been found recently, however, that the matrix configurations
generated by the simulation is singular
in that the submatrices representing the expanding 3D space
have only two large eigenvalues associated with 
the Pauli matrices \cite{workinprog}.
This problem has been attributed to the aforementioned
approximation used to avoid the sign problem since 
the function obtained after integrating out the scale factor is
actually complex-valued, and the effect of the phase is 
not taken into account.
It was realized that the approximation actually amounts
to replacing the phase factor $\ee^{iS_{\rm b}}$ 
by a positive definite weight $\ee^{c S_{\rm b}}$ with 
some constant $c >0$.
This new interpretation of the simulation provides
clear understanding of the observed
Pauli-matrix structure and the (3+1)D expanding behavior.
It has also been argued that
a regular space-time may be obtained if the phase factor $\ee^{iS_{\rm b}}$ 
is used correctly.
This is a very nontrivial issue, however, since losing 
the Pauli-matrix structure may also imply 
losing the (3+1)D expanding behavior at the same time.
%


In this paper we address this issue by using the CLM
to solve the sign problem
instead of using the aforementioned approximation.
Note that the Lorentzian type IIB matrix model
needs to be regularized in some way or another 
because the phase factor $\ee^{iS_{\rm b}}$
in the partition function cannot suppress
the contribution from the bosonic matrices with arbitrary 
large elements.\footnote{This situation is in sharp contrast to the 
Euclidean version \cite{Krauth:1998xh,Austing:2001pk},
in which the phase factor $\ee^{iS_{\rm b}}$ is 
replaced by $\ee^{-S_{\rm b}^{({\rm E})}}$ where 
$S_{\rm b}^{({\rm E})}$ is a real non-negative quantity.}
Here we use the infrared cutoffs on both the spatial and temporal
matrices analogous to the ones used in the previous work \cite{Kim:2011cr}.
We also find it useful to introduce
two deformation parameters $(s,k)$,
which correspond to the Wick rotations
on the worldsheet and in the target space, respectively.
These parameters enable us to interpolate between the Lorentzian version
$(s,k)=(0,0)$ 
and the Euclidean version $(s,k)=(1,1)$ .

First we focus on
$(s,k)=(-1,0)$ 
in the deformation parameter space, where we do not have 
the sign problem. In fact, this case corresponds to 
the approximate model investigated in our previous work.
We observe the emergence of 
(3+1)D expanding space-time with the Pauli-matrix structure.
Then we tune the worldsheet deformation parameter $s$ 
close to that for the Lorentzian model ($s=0$)
keeping the target space deformation parameter $k$ in such a way that
the space-time noncommutativity is minimized.
There, we find it possible to obtain
a smoother space-time structure
without losing the (3+1)D expanding behavior.
The deviation from the Pauli-matrix structure was not seen 
for the matrix size $N \le 64$ within the parameter region 
that can be explored by the CLM, 
and it becomes more prominent as we increase $N$ from 128 to 192.
We consider that the two deformation parameters $s$ and $k$ should
be tuned eventually to $(s,k)=(0,0)$
in the large-$N$ limit.
Whether a smooth classical space-time picture appears
in that limit at sufficiently late time is an important open question,
which can be answered
along the line of this research.

The rest of this paper is organized as follows. 
In section \ref{sec:definition} we
define the Lorentzian type IIB matrix model
and introduce the infrared cutoffs as well as the
two deformation parameters $s$ and $k$.
In section \ref{sec:CLM} we discuss how we apply the CLM
to the Lorentzian type IIB matrix model.
In section \ref{sec:expanding} we focus on
$(s,k)=(-1,0)$ in the deformation parameter space, 
which corresponds to 
the approximate model investigated in the previous work.
Indeed we observe the emergence of 
(3+1)D expanding space-time with the Pauli-matrix structure.
In section \ref{sec:departure} we show our results for
the worldsheet deformation parameter $s$ 
close to that for the Lorentzian model ($s=0$)
with the target space deformation parameter $k$ 
chosen in such a way that
the space-time noncommutativity is minimized.
We observe a clear departure from the Pauli-matrix structure,
while the (3+1)D expanding behavior is still being observed.
Section \ref{sec:summary} is devoted to a summary and discussions.

\section{Definition of the Lorentzian type IIB matrix model}
\label{sec:definition}

The action of the Lorentzian type IIB matrix 
model
is given by \cite{Ishibashi:1996xs}
\begin{eqnarray}
S & = & S_{{\rm b}}+S_{{\rm f}} \ ,
\label{eq:S_likkt}\\
S_{{\rm b}} & = & 
- \frac{1}{4g^{2}}{\rm Tr}
\left(\left[A_{\mu},A_{\nu}\right]
\left[A^{\mu},A^{\nu}\right]\right) \ ,
\label{eq:Sb}\\
S_{{\rm f}} & = & 
-\frac{1}{2g^{2}}{\rm Tr}
\left(\Psi_{\alpha}\left(\mathcal{C}
\Gamma^{\mu}\right)_{\alpha\beta}
\left[A_{\mu},\Psi_{\beta}\right]\right) \ ,
\label{eq:Sf-1}
\end{eqnarray}
where the bosonic variables
$A_{\mu}$ $\left(\mu=0,\ldots,9\right)$
and the fermionic variables
$\Psi_{\alpha}$ $\left(\alpha=1,\ldots,16\right)$
are $N\times N$ Hermitian matrices. 
$\Gamma^{\mu}$ are 10D gamma-matrices
after the Weyl projection and $\mathcal{C}$ is the charge conjugation
matrix. The ``coupling constant'' $g$ is merely a scale parameter
in this model since it can be absorbed by rescaling $A_{\mu}$ and
$\Psi$ appropriately. 
The indices $\mu$ and $\nu$
are contracted using the Lorentzian metric 
$\eta_{\mu\nu}={\rm diag}\left(-1,1,\ldots,1\right)$.
The Euclidean version can be obtained by making
a ``Wick rotation'' $A_0 = i A_{10}$, where $A_{10}$ 
is Hermitian.

The partition function 
for the Lorentzian version
is proposed in ref.~\cite{Kim:2011cr} as 
\begin{equation}
Z=\int dAd\Psi\, e^{iS}
\label{Z-Likkt1}
\end{equation}
with the action \eqref{eq:S_likkt}. 
The ``$i$'' in front of the action
is motivated from the fact that
the string worldsheet metric should also have 
a Lorentzian signature. 
By integrating out the fermionic matrices,
we obtain the Pfaffian 
\begin{equation}
\int d\Psi\, e^{iS_{{\rm f}}} =
{\rm Pf}\mathcal{M}\left(A\right)  \ ,
\end{equation}
which is real unlike in the Euclidean case \cite{Anagnostopoulos:2013xga}.
Note also that the bosonic action \eqref{eq:Sb}
can be written as 
\begin{eqnarray}
S_{\rm b}  =  
\frac{1}{4g^{2}}{\rm Tr}\left(F_{\mu\nu}F^{\mu\nu}\right)
 =  \frac{1}{4g^{2}}
\left\{ {\rm -2Tr}\left(F_{0i}\right)^{2}+
{\rm Tr}\left(F_{ij}\right)^{2}\right\} \ ,
\label{decomp-Sb}
\end{eqnarray}
where we have introduced
the Hermitian matrices $F_{\mu\nu}=i\left[A_{\mu},A_{\nu}\right]$.
Since the two terms
in the last expression
have opposite signs,
$S_{{\rm b}}$ is not positive semi-definite,
and it is not bounded from below.

In order to make the partition function \eqref{Z-Likkt1} finite,
we need to introduce infrared cutoffs 
in both the temporal and spatial directions, 
for instance, as
\begin{eqnarray}
\frac{1}{N}{\rm Tr}\left(A_{0}\right)^{2} 
& \le & \kappa
L^2
\ ,\label{eq:t_cutoff}\\
\frac{1}{N}{\rm Tr}\left(A_{i}\right)^{2} 
& \le & L^{2} \ .
\label{eq:s_cutoff}
\end{eqnarray}

We can use
the ${\rm SU}\left(N\right)$ symmetry of the model
to bring the temporal matrix $A_{0}$ into the diagonal form
\begin{equation}
A_{0}={\rm diag}\left(\alpha_{1},\ldots,\alpha_{N}\right)\ ,
\quad \quad
{\rm where~} \alpha_{1}<\cdots<\alpha_{N} \ .
\label{eq:diagonal gauge}
\end{equation}
By ``fixing the gauge'' in this way,
we can rewrite the partition function (\ref{Z-Likkt1}) as
\beqa
\label{gauge-fixing}
Z &=& \int  \prod_{a=1}^{N}d\alpha_{a}\,
\Delta (\alpha)^2  \int dA_i \, e^{iS_{\rm b}} 
{\rm Pf}\mathcal{M}\left(A\right) \ , 
\\
\Delta (\alpha) &\equiv &
\prod_{a>b}^{N}
\left(\alpha_{a}-\alpha_{b}\right) \ ,
\label{A0diag}
\eeqa
where $\Delta(\alpha)$ is the van der Monde determinant.
The factor $\Delta (\alpha)^2$ 
in (\ref{gauge-fixing})
appears from the Fadeev-Popov procedure
for the gauge fixing, and it acts as a repulsive potential 
between the eigenvalues $\alpha_i$ of $A_0$.

We can extract a time-evolution
from configurations generated by simulating (\ref{gauge-fixing}).
A crucial observation is that 
the spatial matrices $A_{i}$ have
a band-diagonal structure
in the SU($N$) basis in which $A_{0}$ 
has the diagonal form (\ref{eq:diagonal gauge}).
More precisely, there exists some integer $n$ such that
the elements of spatial matrices
$\left(A_{i}\right)_{ab}$ for $\left|a-b\right|>n$ are 
much smaller than those for $\left|a-b\right|\leq n$.
Based on this observation,
we may naturally consider $n\times n$ submatrices of $A_i$ defined as
\begin{equation}
\left(\bar{A}_{i}\right)_{IJ}\left(t\right)
\equiv\left(A_{i}\right)_{\nu+I,\nu+J} \ ,
\label{eq:def_abar}
\end{equation}
where $I,J=1,\ldots , n$, $\nu=0,1,\ldots , N-n$,
and $t$ is defined by 
\begin{equation}
t=\frac{1}{n}\sum_{I=1}^{n}\alpha_{\nu+I} \ .
\label{eq:def_t}
\end{equation}
We interpret the $\bar{A}_i(t)$ 
as representing the state of the universe at time $t$.


Using $\bar{A}_{i}(t)$,
we can define, for example,
the extent of space at time $t$ as 
\begin{equation}
R^{2}\left(t\right)=
\left\langle \frac{1}{n}{\rm tr}\sum_{i}
\left(\bar{A}_{i}\left(t\right)\right)^{2}\right\rangle \ ,
\label{eq:def_rsq}
\end{equation}
where the symbol ${\rm tr}$ represents
a trace over the $n\times n$ submatrix.
We also define
the ``moment of inertia tensor'' 
\begin{equation}
T_{ij}\left(t\right)
=\frac{1}{n}{\rm tr}
\Big(\bar{A}_{i}(t) \bar{A}_{j}(t)\Big) \ ,
\label{eq:def_tij}
\end{equation}
which is a $9\times9$ real symmetric matrix. 
The eigenvalues of $T_{ij}\left(t\right)$,
which we denote by $\lambda_{i}\left(t\right)$ with the order
\begin{equation}
\lambda_{1}\left(t\right)>\lambda_{2}
\left(t\right)>\cdots>\lambda_{9}\left(t\right)
\end{equation}
represent the spatial extent in each of 
the nine directions at time $t$.
Note that the expectation values 
$\left\langle \lambda_{i}\left(t\right)\right\rangle $
tend to be equal in the large-$N$ limit if the SO(9) symmetry is
not spontaneously broken. 
This is the case at early times of the time-evolution.
After a critical time $t_{{\rm c}}$, on the other hand,
it was found \cite{Kim:2011cr}
that the three largest eigenvalues
$\left\langle \lambda_{i}\left(t\right)\right\rangle$ 
($i=1$, $2$, $3$)
become significantly larger than the rest,
which implies that
the SO(9) symmetry is spontaneously broken down to SO(3).

Here we introduce two deformation parameters $s$ and $k$,
which correspond to Wick rotations on the worldsheet and 
in the target space, respectively.
Let us introduce $\tilde{S}=- i S_{\rm b}$ so that
the factor $e^{iS_{\rm b}}$
in the partition function (\ref{gauge-fixing}) 
is rewritten as $e^{- \tilde{S}}$.
We introduce the first parameter $s$ ($-1 \le s \le 1$)
corresponding to the Wick rotation on the worldsheet as
\begin{equation}
\tilde{S} = - i N \beta \, \ee^{ i s \pi/ 2}
\left\{ 
- \frac{1}{2} \Tr (F_{0i})^2 
+ \frac{1}{4} \Tr (F_{ij})^2 
\right\} \ ,
\label{sdef-action}
\end{equation}
where
$\beta = \frac{1}{g^2 N}$.
The second parameter $k$ ($0 \le k \le 1$)
corresponding to the Wick rotation in the target space
can be introduced by the replacement $A_0 \mapsto  \ee^{- i k \pi /2} A_0$.
The action (\ref{sdef-action}) becomes
\begin{equation}
\tilde{S} = - i N \beta \,
\ee^{ i s \pi/ 2}
\left\{ 
- \frac{1}{2} \ee^{- i k \pi} \Tr (F_{0i})^2 
+ \frac{1}{4} \Tr (F_{ij})^2 
\right\}  \ ,
\label{sdef-action2}
\end{equation}
and the ${\rm Pf}\mathcal{M}\left(A\right)$ in
(\ref{gauge-fixing})
should be replaced by 
${\rm Pf}\mathcal{M}(\ee^{- i k \pi /2} A_0, A_i)$.
The Lorentzian model is retrieved at $(s,k)=(0,0)$,
whereas the Euclidean model corresponds to setting $(s,k)=(1,1)$.

Note that 
the coefficient of the first term in
(\ref{sdef-action2})
can be made
real non-negative by choosing the parameters so that
$ i \ee^{ i s \pi/ 2} \ee^{- i k \pi} = 1 $,
which implies $k=(1+s)/2$.
For this choice, the bosonic action 
is most effective in
minimizing the noncommutativity between
the spatial matrices $A_i$ and the temporal matrix $A_0$.
For $0 \le k < s/2$, on the other hand, 
the real part of the coefficient becomes negative,
which favors maximum noncommutativity between $A_i$ and $A_0$.
As a result, the eigenvalues of $A_0$ 
lump up into two clusters separated from each other,
and we cannot obtain a continuous time.
The Lorentzian model $(s,k)=(0,0)$ lies on the boundary of this
unphysical region.
In this work, we keep away from this region by
restricting ourselves to the cases satisfying $k=(1+s)/2$.

Taking into account the infrared cutoffs 
(\ref{eq:t_cutoff}) and (\ref{eq:s_cutoff}),
we arrive at the partition function
\beqa
Z &=& 
\int  \prod_{a=1}^{N}d\alpha_{a}\,
\Delta (\alpha)^2  \int dA_i \, e^{- \tilde{S}} 
{\rm Pf}\mathcal{M}(\ee^{- i k \pi /2} A_0, A_i) 
\nn
\\
&~& 
\times  \theta\left(\kappa L^2 -  \frac{1}{N}{\rm Tr}(A_{0})^{2}\right)
\theta\left(L^2 -  \frac{1}{N}{\rm Tr}(A_{i})^{2} \right)
\ ,
\label{Z-Likkt-n1-cutoff}
\eeqa
where $\theta(x)$ is the Heaviside step function
and $\tilde{S}$ is given by (\ref{sdef-action2}).
By rescaling $A_\mu \mapsto L A_\mu$ 
and $\beta \mapsto L^{-4} A_\mu$,
we can set $L=1$ without loss of generality.

\section{Applying the CLM to the Lorentzian model}
\label{sec:CLM}

We apply the CLM to the model (\ref{Z-Likkt-n1-cutoff}).
From now on, we omit the Pfaffian
and consider the 
6D version, which consists of $A_0$
and $A_i$ ($i=1,\cdots , 5$), for simplicity.

The first step of the CLM is to complexify the real variables.
As for the spatial matrices $A_i$, 
we simply treat them
as general complex matrices instead of Hermitian matrices.
As for the temporal matrix $A_0$, which is diagonalized as
(\ref{eq:diagonal gauge}),
we have to take into account the ordering of the eigenvalues.
For that purpose,
we make the change of variables as
\begin{alignat}{3}
\alpha_1 = 0 \ , \quad
\alpha_2 = \ee^{\tau_1} \ , \quad
\alpha_3 = \ee^{\tau_1} + \ee^{\tau_2} \ ,
\quad
\cdots  \ , \quad 
\alpha_N = \sum_{a=1}^{N-1} 
\ee^{\tau_a}  
  \label{eq:alpha-tau}
\end{alignat}
so that the ordering is implemented automatically,
and then complexify $\tau_a$ ($a=1 ,\cdots , N-1$).
We have chosen to set $\alpha_1 =0$ using the shift symmetry
$A_0 \mapsto A_0 + {\rm const}. {\bf 1}$ of the action.
In order to respect this symmetry,
we decide to impose the cutoff like (\ref{eq:t_cutoff})
only on the traceless part 
$\tilde{A}_0 = A_0 - \frac{1}{N} \Tr A_0$
in this work. 

The Heaviside function in (\ref{Z-Likkt-n1-cutoff}) 
is difficult to treat in the CLM as it is.
Here we mimic its effect by introducing the potential
\begin{alignat}{3}
S_{\rm pot}
= \frac{1}{p} \gamma_{\rm s} 
\left(\frac{1}{N} \Tr (A_i)^2 - 1 \right)^p 
+ \frac{1}{p} \gamma_{\rm t} 
\left(\frac{1}{N} \Tr (\tilde{A}_0)^2 
- \kappa  \right)^p  \ ,
  \label{eq:pot}
\end{alignat}
where the power $p$ is set to $p=4$ in this work,
and the coefficients $\gamma_{\rm s} $ and $\gamma_{\rm t}$
are chosen to be large enough to make
$\frac{1}{N} \Tr (A_i)^2$ and $\frac{1}{N} \Tr (\tilde{A}_0)^2$
fluctuate around some constants.\footnote{This appears different
from imposing the inequalities (\ref{eq:t_cutoff}) and
(\ref{eq:s_cutoff}), but
the difference is not important since the inequalities are typically
saturated due to entropic effects.}
The effective action then reads
\begin{alignat}{3}
S_{\rm eff}  
&=  N
\beta
 \,  e^{-i \frac{\pi}{2} (1-s)}  \left\{ 
\frac{1}{2} e^{-i k \pi}
\Tr [A_0 , A_i]^2 
 - \frac{1}{4} 
\Tr [A_i , A_j]^2
\right\} \nonumber \\
& + \frac{1}{p} \gamma_{\rm s} 
\left(\frac{1}{N} \Tr (A_i)^2 - 1 \right)^p 
+ \frac{1}{p} \gamma_{\rm t} 
\left(\frac{1}{N} \Tr (\tilde{A}_0)^2 
- \kappa  \right)^p \nonumber \\
&  - 2 \, \log \Delta(\alpha)  
- \sum_{a=1}^{N-1}\tau_a \ ,
\label{eq:eff_action}
\end{alignat}
where the last term comes from the Jacobian associated
with the change of variables (\ref{eq:alpha-tau}).
The complex Langevin equation is given by
\begin{alignat}{3}
\frac{d\tau_a}{dt}&=
 - \frac{\del S_{\rm eff}}{\del \tau_a} + \eta_a(t)  \ ,   \nn \\
\frac{d (A_i)_{ab}}{dt}&=
 - \frac{\del S_{\rm eff}}{\del (A_i)_{ba}} 
+ (\eta_i)_{ab}(t)  \ ,
  \label{eq:cle}
\end{alignat}
where the $\eta_a(t)$ in the first equation 
are random real numbers
obeying the probability distribution
$\exp (- \frac{1}{4}\int dt \sum_a \{ \eta_a(t) \}^2 )$
and the $\eta_i(t)$ in the second equation 
are random Hermitian matrices
obeying the probability distributions 
$\exp (- \frac{1}{4}\int dt \sum_{i} \Tr  \{ \eta_i(t) \}^2 )$.

The expectation values of observables can be calculated by 
defining them holomorphically for complexified $\tau_a$ and $A_i$
and taking an average using the configurations generated by
solving the discretized version of (\ref{eq:cle}) 
for sufficiently long time.
In order for this method to work, the probability distribution
of the drift terms, namely the first terms
on the right-hand side of (\ref{eq:cle}), has to fall off 
exponentially \cite{Nagata:2016vkn}.
We have checked that this criterion is indeed satisfied
for all the values of parameters used in this paper.

\begin{figure}[t]
\centering
\includegraphics[width=7cm]{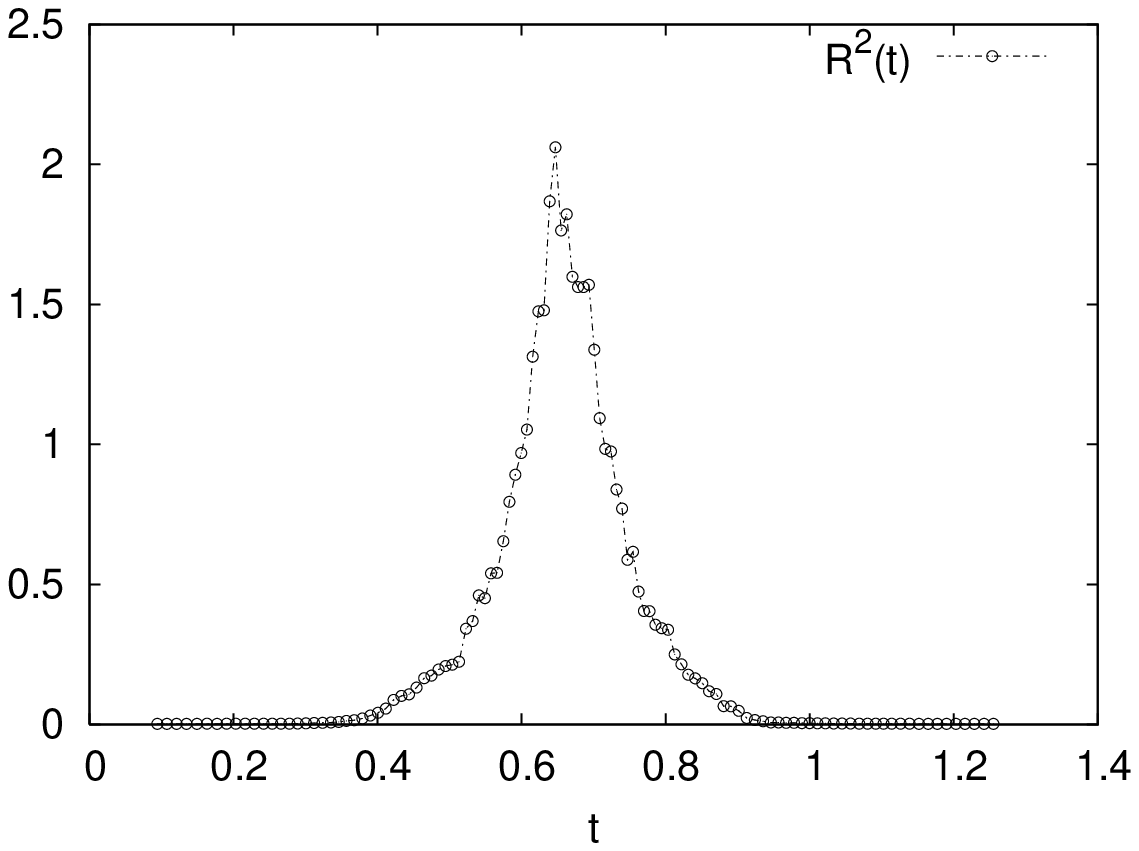}\\
\includegraphics[width=7cm]{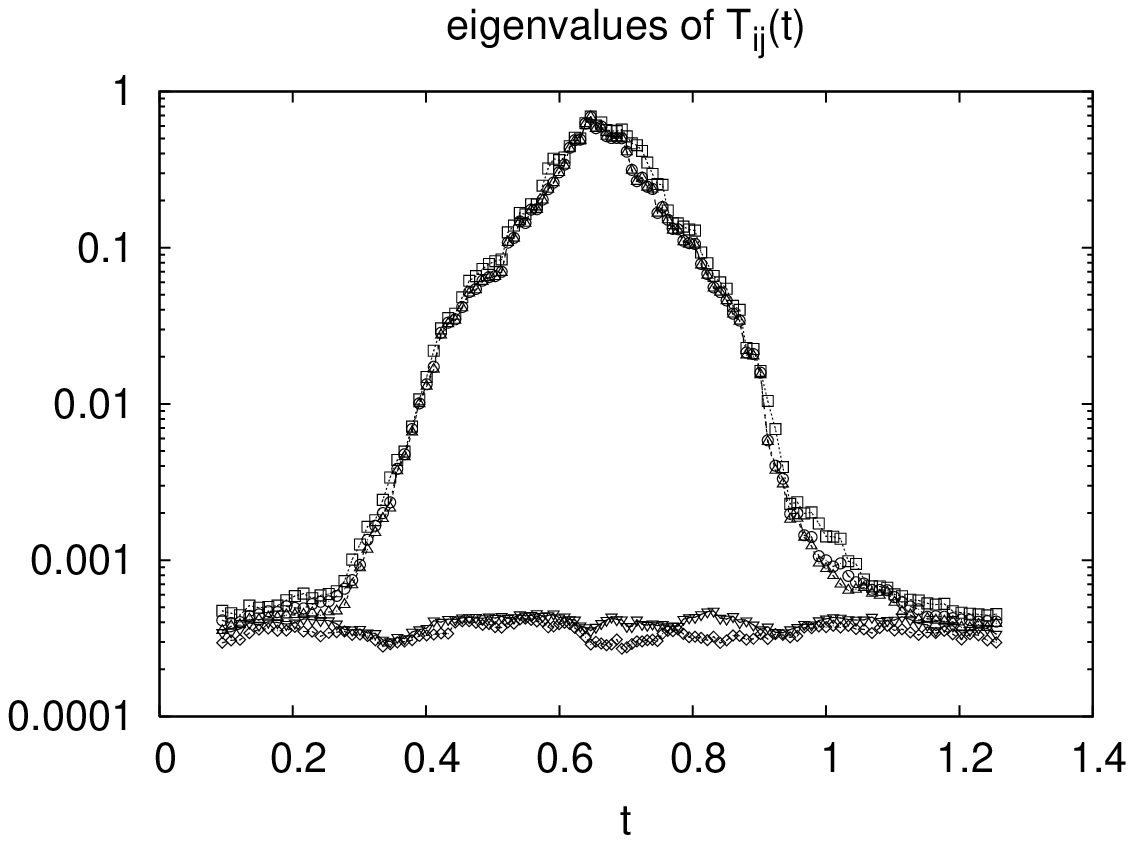}
\includegraphics[width=7cm]{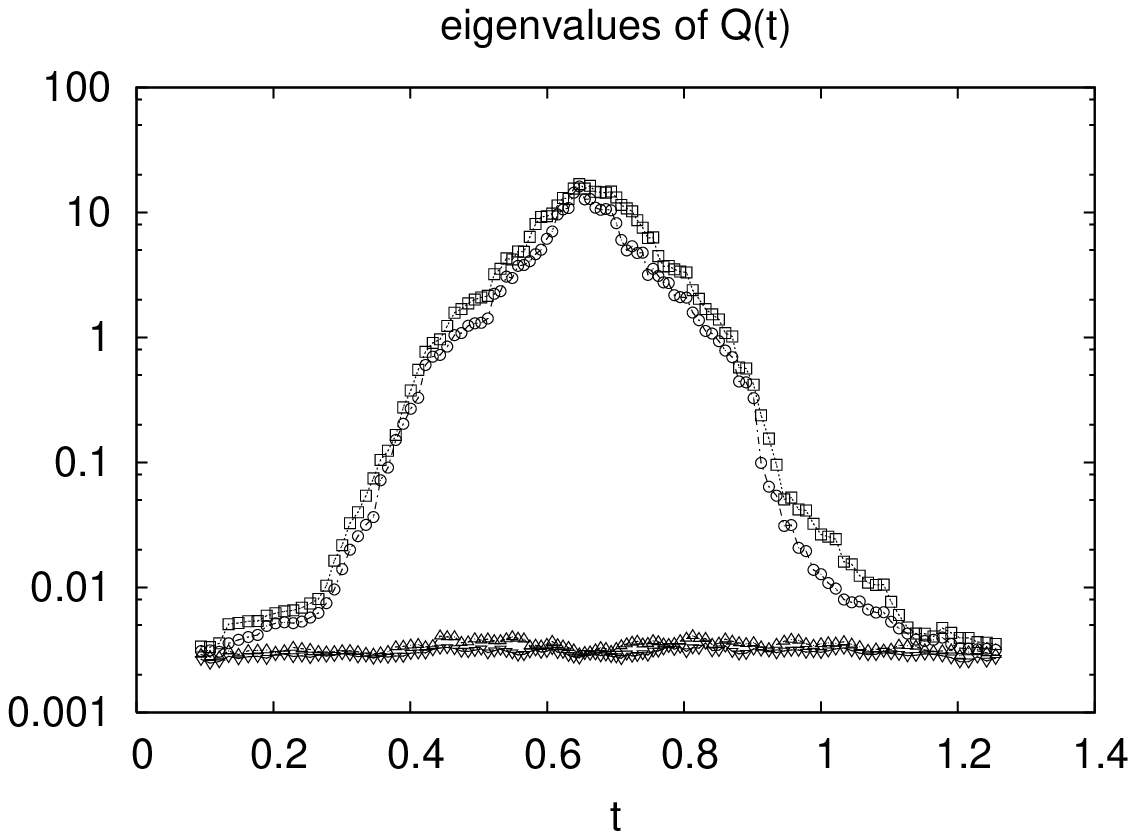}
\caption{Results for $(s,k)=(-1,0)$, $N=128$, $\kappa=0.13$, $\beta=2$,
$n=16$ are shown.
(Top) The extent of space $R^2(t)$ is plotted
against $t$.
(Bottom-Left) The five eigenvalues of the moment of inertia tensor
are plotted against $t$ in the log scale.
(Bottom-Right) The four largest eigenvalues of the matrix $Q(t)$ are plotted
against $t$ in the log scale.}
\label{fig:N=128sdef=-1}
\end{figure}

\section{Emergence of (3+1)D expanding behavior}
\label{sec:expanding}

In this section we consider $(s,k)=(-1,0)$
in the parameter space.
The action is given by
\begin{alignat}{3}
S = N \beta \,
\Big\{ 
- \frac{1}{2}  \Tr [A_0 , A_i]^2 
+ \frac{1}{4} \Tr [A_i , A_j]^2 
\Big\}  \ ,
\label{sdef-action3}
\end{alignat}
which is real, and the CLM reduces 
to the ordinary Langevin method.
The first term in (\ref{sdef-action3})
tries to minimize the space-time noncommutativity,
which has the effects of making
the spatial matrices close to diagonal
in the basis (\ref{eq:diagonal gauge}).
On the other hand, the second term favors maximal noncommutativity
among spatial matrices.

Figure \ref{fig:N=128sdef=-1} 
shows our results\footnote{Here and hence forth,
we plot the results obtained for one thermalized configuration.}
for $N=128$, $\kappa=0.13$, $\beta=2$.
The block size for (\ref{eq:def_abar}) is chosen to be $n=16$.
In the Top panel, we plot the 
extent of space $R^2(t)$ defined
by (\ref{eq:def_rsq}) against $t$.
The result is symmetric under the reflection 
$t - t_{\rm p} \mapsto -(t - t_{\rm p})$, where
$t_{\rm p}$ represents the time at which $R^2(t)$ is peaked,
due to the symmetry of the model 
under $\tilde{A}_0 \mapsto - \tilde{A}_0$.

Next we discuss the SSB of SO(5) symmetry by
considering the moment of inertia tensor (\ref{eq:def_tij}).
In the Bottom-Left panel,
we plot the 
eigenvalues $\lambda_{i}(t)$ of $T_{ij}(t)$, which
shows that only three out of five eigenvalues become large
in the time region around $t=t_{\rm p}$.
This suggests that the rotational SO(5) symmetry 
of the 6D bosonic model is broken down to SO(3) in that time region.
These results are qualitatively the same as
what has been obtained in ref.~\cite{Ito:2015mxa},
which is consistent with the speculation \cite{workinprog}
that the previous simulations correspond to 
the parameter choice $(s,k)=(-1,0)$.

As is known from the previous work \cite{Kim:2011cr},
the time difference between the peak ($t=t_{\rm p}$)
and the critical time at which the SSB occurs
increases in physical units
as we take the large-$N$ limit.
Therefore, the reflection symmetry with respect to $t$
does not necessarily
imply that the Big Crunch occurs in the finite future.

The mechanism of this SSB can be understood as follows \cite{workinprog}.
Since the first term in (\ref{sdef-action3}) favors $A_i$ close to diagonal, 
we may consider the submatrices $\bar{A}_i(t)$ as the effective
degrees of freedom.
The infrared cutoff (\ref{eq:s_cutoff}) fixes
$\Tr \{ \bar{A}_i(t) \} ^2$ to some constant, and 
the second term in (\ref{sdef-action3}) favors maximal noncommutativity
between $\bar{A}_i(t)$.
According to the argument in ref.~\cite{Kim:2011cr}, this leads
to $\bar{A}_i(t)\propto \sigma_i \oplus {\bf 0}_{n-2}$ 
for $i=1,2,3$ and 
$\bar{A}_i(t) = {\bf 0}_{n}$ for $i\ge 4$ up to SO(5) rotations,
where $\sigma_i$ are the Pauli matrices.
In order to confirm this mechanism, we calculate the matrix 
\begin{alignat}{3}
Q(t)=\sum_{i=1}^5 \{ \bar{A}_i (t) \}^2 \ ,
\label{def-Q}
\end{alignat}
and plot the 
four largest eigenvalues of $Q(t)$ in 
Fig.~\ref{fig:N=128sdef=-1} (Bottom-Right).
Indeed we find 
that only two of them are large, while the rest are very small
in the time region in which the SSB occurs.


\begin{figure}[t]
\centering
\includegraphics[width=7cm]{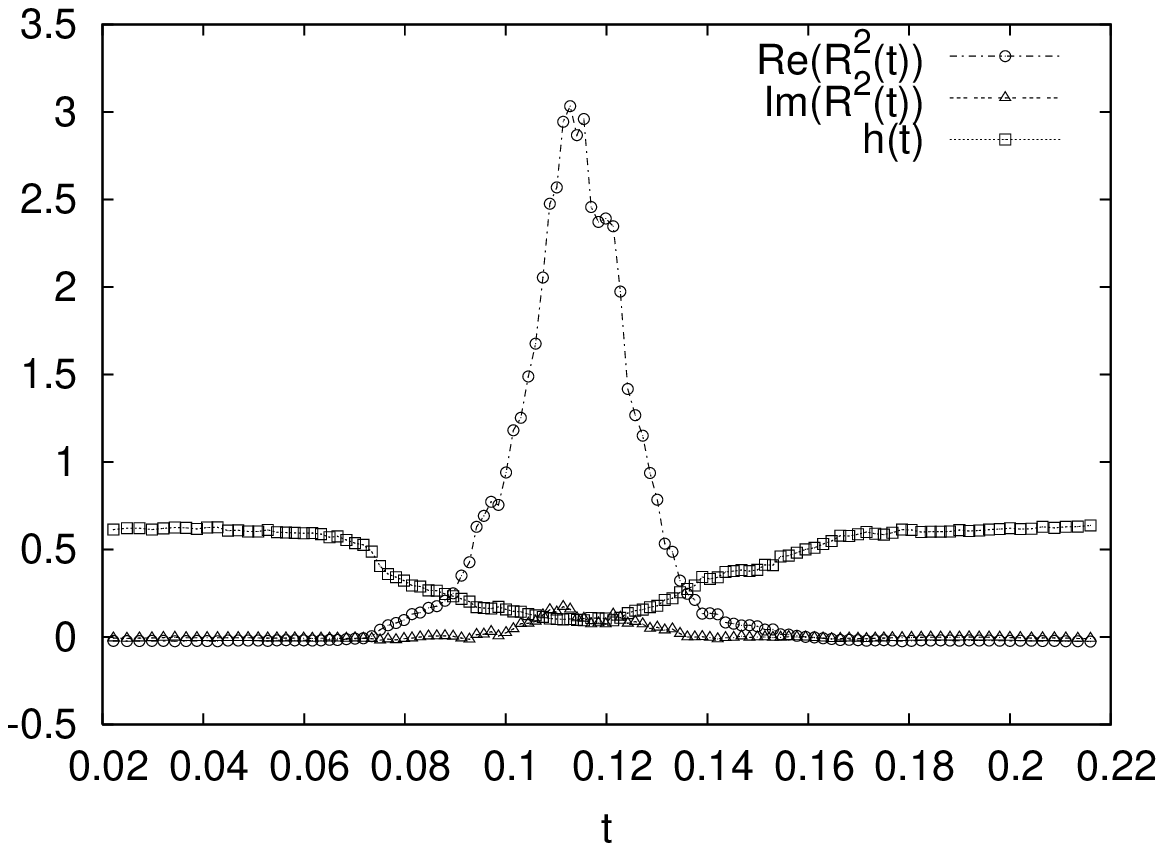}\\
\includegraphics[width=7cm]{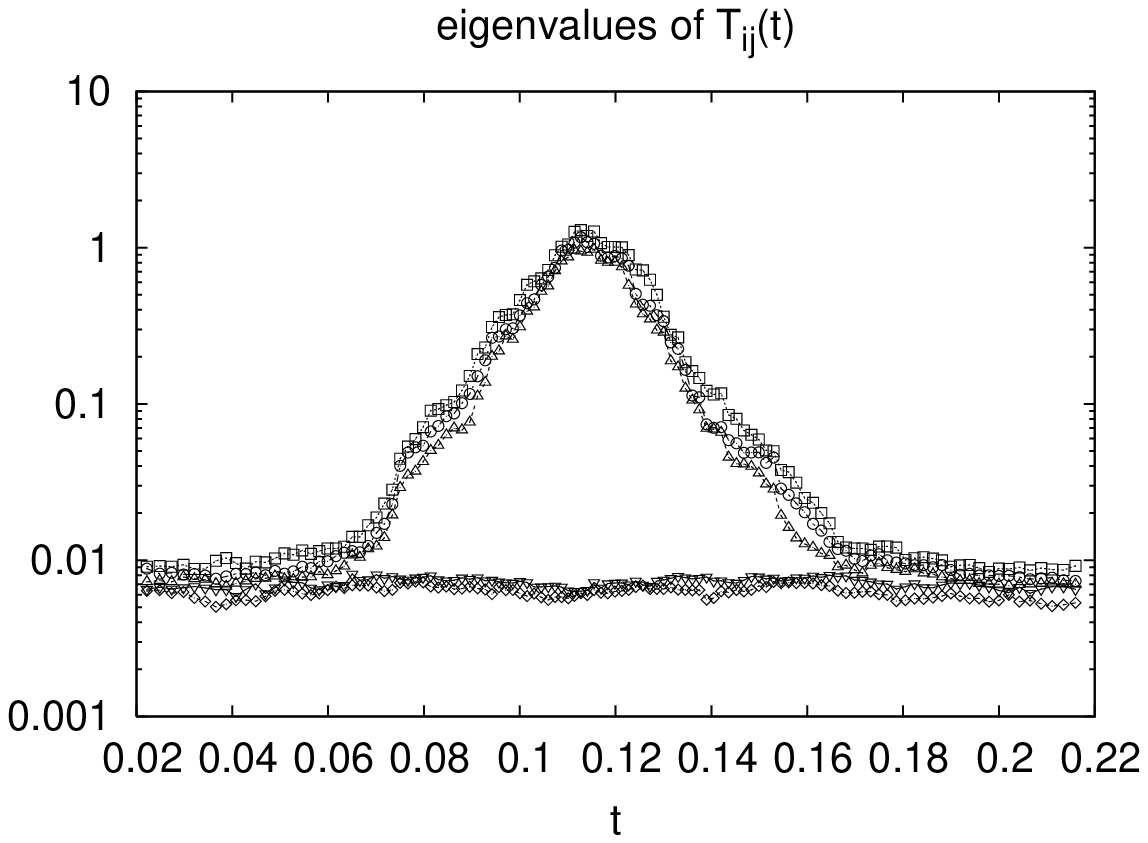}
\includegraphics[width=7cm]{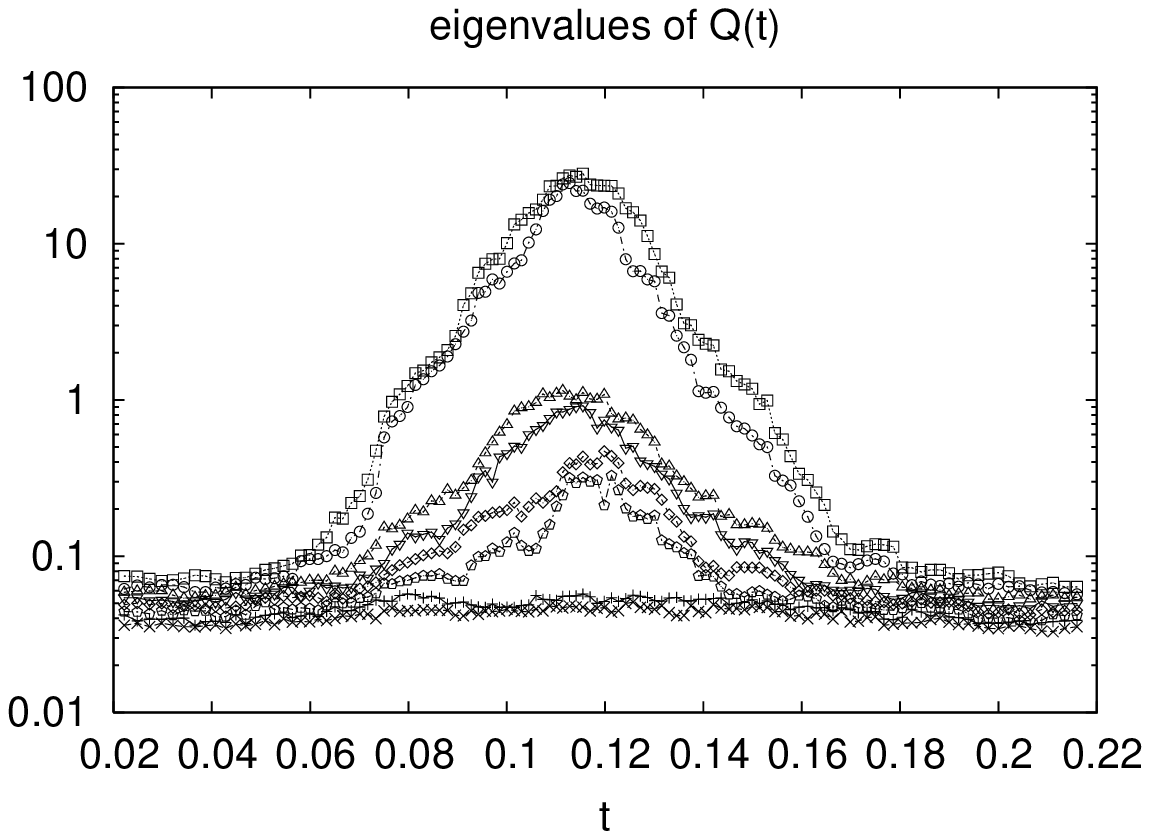}
\caption{Results for $(s,k)=(0.0076,0.5038)$, 
$N=128$, $\kappa=0.0037$, $\beta=32$, $n=16$ are shown.
(Top) The real and imaginary parts of $R^2(t)$ are plotted
against $t$. 
The Hermiticity norm $h(t)$ of the matrix $\bar{A}_i(t)$ is also plotted.
(Bottom-Left) The five eigenvalues of the moment of inertia tensor
are plotted against $t$ in the log scale.
(Bottom-Right) The eight largest eigenvalues of the matrix $Q(t)$ are plotted
against $t$ in the log scale.
}
\label{fig:N=128sdef=0.0076}
\end{figure}

\section{Departure from the Pauli-matrix structure}
\label{sec:departure}

In this section we tune the worldsheet deformation parameter $s$ to
some values near $s=0$, 
which is the target value for the Lorentzian model, 
keeping the target-space deformation parameter $k$ to be
$k=(1+s)/2$, which minimizes the space-time noncommutativity.
The action reads
\begin{alignat}{3}
S = N \beta \,
\Big\{ - \frac{1}{2}  \Tr [A_0 , A_i]^2 
- \ee^{-i\frac{\pi}{2}(1-s)}
\frac{1}{4} \Tr [A_i , A_j]^2 
\Big\}  \ .
\label{sdef-action4}
\end{alignat}
The only difference from (\ref{sdef-action3})
is the second term
with the coefficient $\ee^{-i\frac{\pi}{2}(1-s)}$
whose real part changes its sign at $s=0$.
This implies, in particular, that for $s>0$ 
the second term starts to minimize the noncommutativity
among the spatial matrices.
Therefore, we may anticipate a drastic change of the behavior around $s=0$.
In fact, for the values of $s$ below what is reported below, we
do not see any qualitative difference from the results obtained at $s=-1$.

\begin{figure}[t]
\centering
\includegraphics[width=7cm]{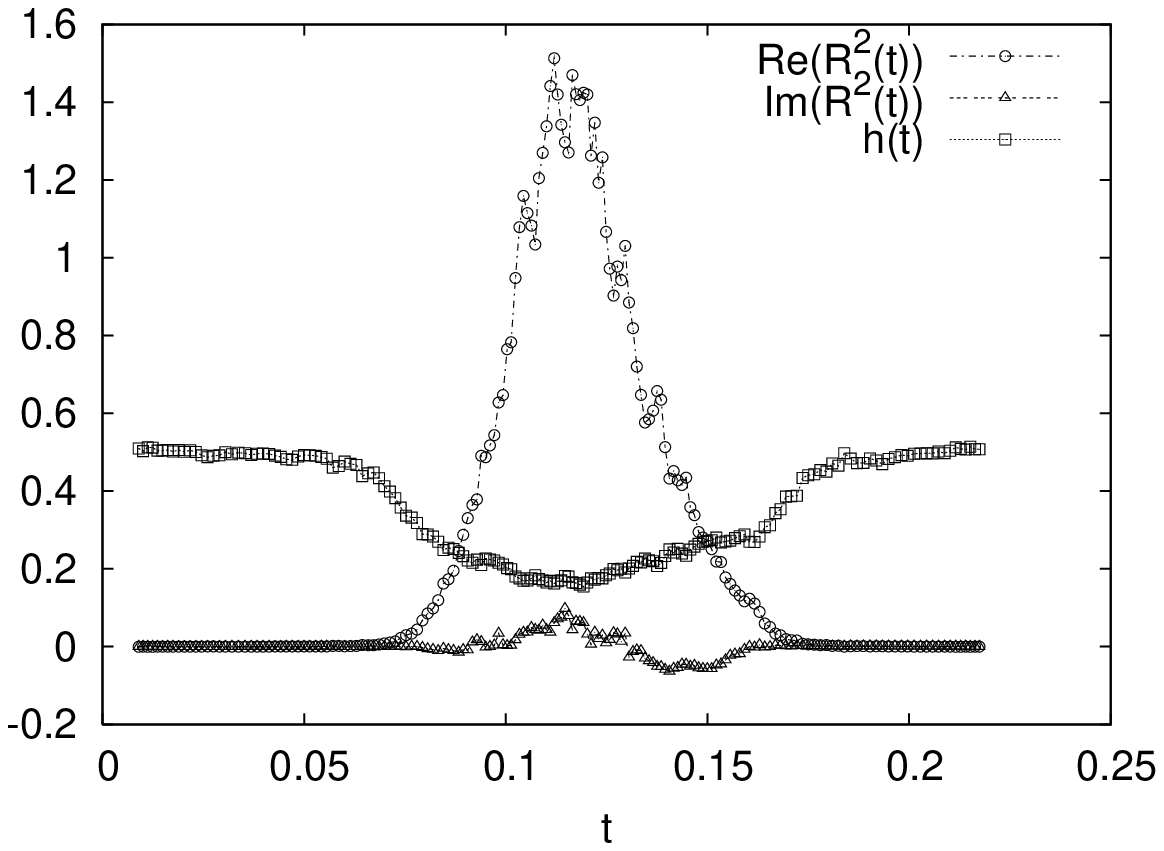}\\
\includegraphics[width=7cm]{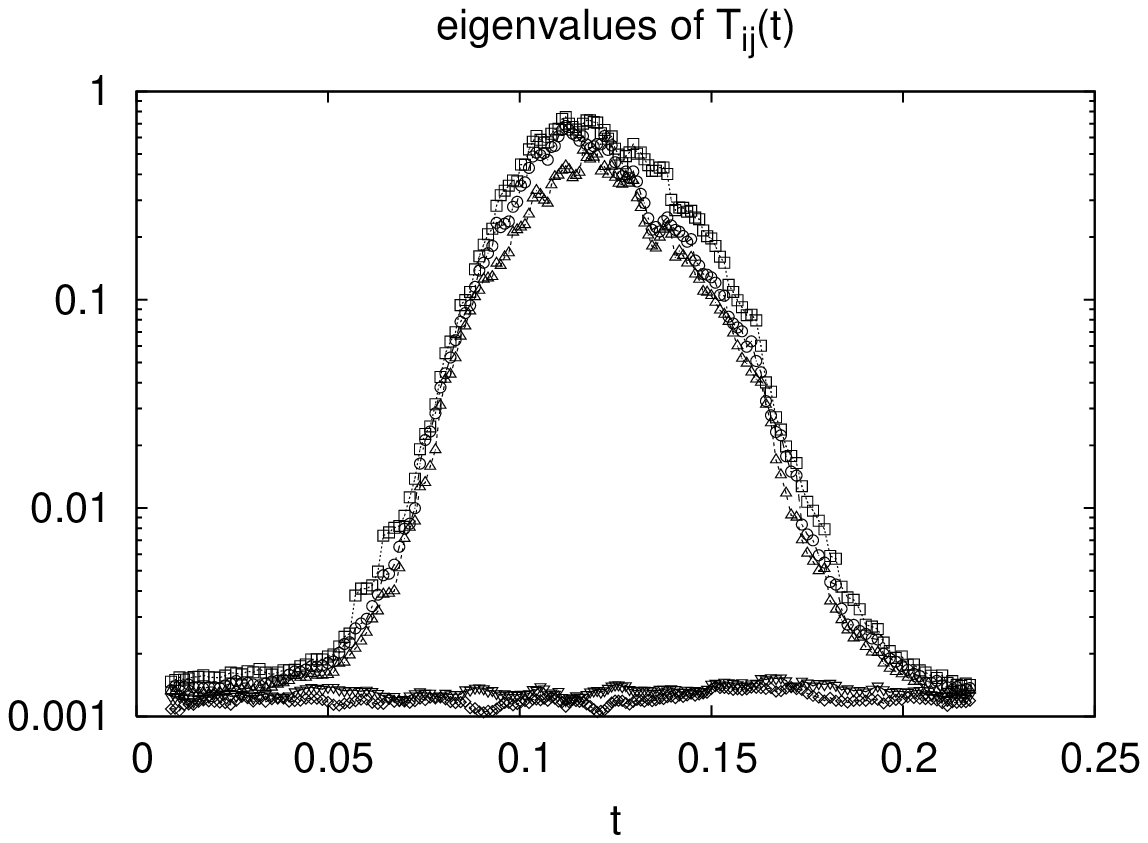}
\includegraphics[width=7cm]{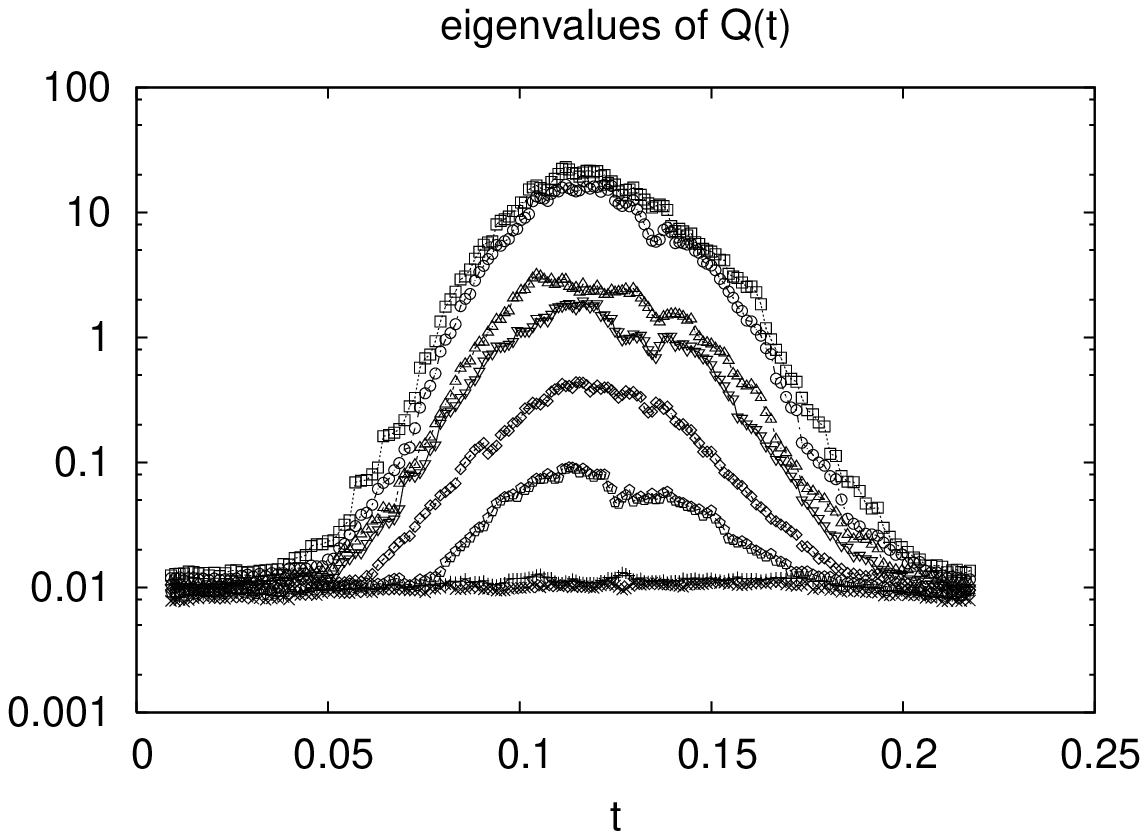}
\caption{Results for $(s,k)=(0.0118,0.5059)$, 
$N=192$, $\kappa=0.0044$, $\beta=64$, $n=24$ are shown.
(Top) The real and imaginary parts of $R^2(t)$ are plotted
against $t$.
The Hermiticity norm $h(t)$ of the matrix $\bar{A}_i(t)$ is also plotted.
(Bottom-Left) The five eigenvalues of the moment of inertia tensor
are plotted against $t$ in the log scale.
(Bottom-Right) The eight largest eigenvalues of the matrix $Q(t)$ are plotted
against $t$ in the log scale.
}
\label{fig:N=192sdef=0.0118}
\end{figure}

Figure \ref{fig:N=128sdef=0.0076}
shows our results
for $N=128$, $\kappa=0.0037$, $\beta=32$, $n=16$
with $(s,k)=(0.0076,0.5038)$.
Unlike the $(s,k)=(-1,0)$ case, 
the action becomes complex 
for $s>-1$ in general.
Therefore, the quantity such as $R^2(t)$ defined in (\ref{eq:def_rsq})
is not guaranteed to be real 
positive.\footnote{Similarly, the time $t$ defined by (\ref{eq:def_t})
is not guaranteed to be real.
However, it turns out to be close to real for the configurations generated
by the complex Langevin method.
We therefore neglect the small imaginary part of $t$ in making the plots in
figures \ref{fig:N=128sdef=0.0076} and \ref{fig:N=192sdef=0.0118}.}
In the Top panel,
we plot the real and imaginary parts of $R^2(t)$.
We find that $R^2(t)$ is dominated by the real part
near the peak.

Let us also take a look at the ``Hermiticity norm''
for $\bar{A}_{i}(t)$ defined by
\begin{alignat}{3}
h(t) =
\frac{- \Tr 
(\bar{A}_i(t) - \bar{A}_i(t)^\dag)^2}
{4 \, \Tr (\bar{A}_i(t) ^\dag \bar{A}_i(t))}
\label{hermiticity-norm} \ ,
\end{alignat}
using a configuration generated by the simulation.
The result is plotted in the 
Top panel
as well.
Note that $h(t)=0$ implies
that the matrices $\bar{A}_i(t)$ are all Hermitian,
while $h(t)=1$ implies
that they are all anti-Hermitian.
We find that $h(t)$ is small and hence
the $\bar{A}_{i}(t)$ are close to Hermitian
near the peak,
which is consistent with 
our observation that $R^2(t)$ is dominated by the real part
in this region.
This property supports our previous 
speculation \cite{Kim:2011ts,Kim:2012mw}
that some classical solution, which is typically 
represented by a real configuration,
dominates the path integral in the time region near the peak
due to the expansion of space.

In the Bottom panels, 
we plot the 
same quantities\footnote{In fact,
it is not straightforward to calculate the expectation values of
the eigenvalues of $T_{ij}(t)$ and $Q(t)$ in the CLM
respecting holomorphicity because of their multi-valuedness.
Here we simply evaluate $T_{ij}(t)$ and $Q(t)$
using the Hermitian part of $\bar{A}_{i}(t)$
from one configuration
generated by the complex Langevin simulation,
and plot their eigenvalues.}
as in Fig.~\ref{fig:N=128sdef=-1} (Bottom).
From the left panel, we observe that (3+1)D expanding behavior
persists even at slightly positive $s$, while the right panel reveals
a clear departure from the Pauli-matrix structure.

%


We perform a similar analysis with the matrix size $N$ increased from 
$N=128$ to $N=192$.
Figure \ref{fig:N=192sdef=0.0118}
shows our results
for $(s,k)=(0.0118,0.5059)$, $N=192$, $\kappa=0.0044$, $\beta=64$, $n=24$.
In the Top panel,
we plot 
the real and imaginary parts of $R^2(t)$
and the Hermiticity norm $h(t)$ defined by 
(\ref{hermiticity-norm}).
As we have seen in Fig.~\ref{fig:N=128sdef=0.0076} (Top),
the spatial matrices $\bar{A}_i(t)$ are close to Hermitian
near the peak of $R^2(t)$, which suggests 
that the behavior in this region
is semi-classical.
In the Bottom panels,
we plot the same quantities as in Fig.~\ref{fig:N=128sdef=0.0076} (Bottom).
We find that the departure from the Pauli-matrix structure
is even more pronounced\footnote{In fact, for both $N=128$ and $N=192$,
we observed the departure from the Pauli-matrix structure
only for slightly positive $s$, where 
the $\Tr (F_{ij})^2$ term starts to favor commutative spatial matrices $A_i$.
This may be due to some quantum effects.}, 
while the (3+1)D expanding behavior is kept intact.

We have also investigated the model with $N<128$.
For $N=32$, $64$, while the results at $(s,k)=(-1,0)$ are similar
to those for $N=128$,
the departure from the Pauli-matrix structure does not show up 
at all even at $s \sim 0$.
As we increase $s$ further with $k=(1+s)/2$,
the Hermiticity
of the configurations is completely lost, and the criterion for 
justifying the CLM is found to be violated.
Thus the use of large values of $N$ seems to be crucial
in investigating the model near the target values $(s,k)=(0,0)$.


\section{Summary and discussions}
\label{sec:summary}

The Lorentzian type IIB matrix model is a promising candidate
for a nonperturbative formulation of superstring theory.
Monte Carlo studies of this model is extremely hard due to 
the sign problem caused by 
the phase factor $\ee^{iS_{\rm b}}$ in the partition function.
Previous work avoided this problem by integrating out the scale factor
of the bosonic matrices and using an approximation.
However, it was noticed recently that
this approximation actually amounts to 
replacing $\ee^{iS_{\rm b}}$ by $\ee^{c S_{\rm b}}$ for some $c>0$.
This suggests the importance of studying the model 
without such an approximation.

In this paper we 
have investigated the space-time structure
based on the complex Langevin simulation 
of the (5+1)D bosonic version of the model with the deformation parameters
$s$ and $k$ corresponding to the Wick rotations on the worldsheet and 
in the target space, respectively.
%
The original model corresponds to $(s,k)=(0,0)$,
whereas our previous simulations were speculated to
correspond to the $(s,k)=(-1,0)$ case \cite{workinprog}.
Our results for $(s,k)=(-1,0)$ indeed
reproduced the (3+1)D expanding behavior with 
the Pauli-matrix structure as expected.
Then we tuned the parameter $s$ towards 
the region $s \sim 0$
restricting ourselves to $k=(1+s)/2$ in order to stabilize
our simulation.
The results indeed showed a clear departure from
the Pauli-matrix structure, while the (3+1)D expanding behavior
is kept intact.
The spatial matrices turn out to be close to Hermitian near the peak 
of the spatial extent $R^2(t)$ even for $s\sim 0$,
which confirms our expectation \cite{Kim:2011ts,Kim:2012mw}
that some classical solution dominates
at late times.

The appearance of the Pauli-matrix structure
for $(s,k)=(-1,0)$
is due to the $\Tr (F_{ij})^2$ term in the action, which tries
to make the spatial matrices $A_i$ maximally noncommutative.
The situation changes drastically around $s=0$, where
the coefficient of the $\Tr (F_{ij})^2$ term becomes
pure imaginary.
On the other hand, there are infinitely many classical 
solutions \cite{Kim:2011ts,Kim:2012mw},
which have (3+1)D expanding behavior without
the Pauli-matrix structure.
(See also refs.\cite{Chaney:2015ktw,Chaney:2015mfa,Chaney:2016npa,Stern:2018wud,Steinacker:2017vqw,Steinacker:2017bhb} for related work.)
We therefore consider it possible that
the space-time structure becomes smooth
without losing the (3+1)D expanding behavior 
in the large-$N$ limit.
The departure from the Pauli-matrix structure
observed at $s\sim 0$ supports this possibility.

Some future directions are in order.
The most important thing to do is to repeat the same analysis
with increased matrix size $N$.
In particular, we need to confirm the appearance of 
a smooth space-time at $(s,k)\sim (0,0)$.
While this issue may not depend much on 
the effects of the fermionic matrices,
it would be certainly desirable to include them eventually.
Unfortunately, this is not straightforward
since the complex Langevin method may suffer from the singular-drift
problem due to the near-zero eigenvalues of the Dirac operator.
The deformation technique \cite{Ito:2016efb}
used successfully in studying
the Euclidean version \cite{Anagnostopoulos:2017gos}
is worth trying, though.
We consider that the dominance of classical solutions at late times
\cite{Kim:2011ts,Kim:2012mw}
supported by our results is important because it enables us to understand
possible late-time behaviors of this model
by solving classical equations of motion.
For instance, we may try to find classical 
solutions \cite{Chatzistavrakidis:2011gs,Nishimura:2013moa,Steinacker:2014fja,Aoki:2014cya},
which can accommodate Standard Model particles as excitations around them.
Work in this direction is ongoing \cite{on-going}.

We hope that the simulation method 
as well as the obtained results
discussed in this paper
is useful in 
understanding the dynamics of the Lorentzian 
type IIB matrix model further.


\section*{Acknowledgements}

The authors would like to thank 
K.N.~Anagnostopoulos,
T.~Aoki, Y.~Asano, T.~Azuma,
M.~Hirasawa, Y.~Ito,
H.~Kawai and H.~Steinacker for valuable discussions.
Computation was carried out
on PC clusters at KEK and XC40 at YITP in Kyoto University.
J.~N.\ and A.~T.\ were supported in part by Grant-in-Aid 
for Scientific Research (No.\ 16H03988 and 18K03614, respectively)
from Japan Society for the Promotion of Science. 

\appendix

\bibliographystyle{JHEP}
\bibliography{cle-lorentz_ref}

\providecommand{\href}[2]{#2}\begingroup\raggedright\begin{thebibliography}{10}

\bibitem{Wilson:1974sk}
K.~G. Wilson, {\it {Confinement of Quarks}},  {\em Phys. Rev.} {\bf D10} (1974)
  2445--2459.

\bibitem{Kim:2011cr}
S.-W. Kim, J.~Nishimura, and A.~Tsuchiya, {\it {Expanding (3+1)-dimensional
  universe from a Lorentzian matrix model for superstring theory in
  (9+1)-dimensions}},  {\em Phys. Rev. Lett.} {\bf 108} (2012) 011601,
  [\href{http://arxiv.org/abs/1108.1540}{{\tt arXiv:1108.1540}}].

\bibitem{Ishibashi:1996xs}
N.~Ishibashi, H.~Kawai, Y.~Kitazawa, and A.~Tsuchiya, {\it {A Large N reduced
  model as superstring}},  {\em Nucl. Phys.} {\bf B498} (1997) 467--491,
  [\href{http://arxiv.org/abs/hep-th/9612115}{{\tt hep-th/9612115}}].

\bibitem{Parisi:1984cs}
G.~Parisi, {\it {On Complex Probabilities}},  {\em Phys. Lett.} {\bf B131}
  (1983) 393--395.

\bibitem{Klauder:1983sp}
J.~R. Klauder, {\it {Coherent State Langevin Equations for Canonical Quantum
  Systems With Applications to the Quantized Hall Effect}},  {\em Phys. Rev.}
  {\bf A29} (1984) 2036--2047.

\bibitem{Aarts:2009dg}
G.~Aarts, F.~A. James, E.~Seiler, and I.-O. Stamatescu, {\it {Adaptive stepsize
  and instabilities in complex Langevin dynamics}},  {\em Phys. Lett.} {\bf
  B687} (2010) 154--159, [\href{http://arxiv.org/abs/0912.0617}{{\tt
  arXiv:0912.0617}}].

\bibitem{Aarts:2009uq}
G.~Aarts, E.~Seiler, and I.-O. Stamatescu, {\it {The Complex Langevin method:
  When can it be trusted?}},  {\em Phys. Rev.} {\bf D81} (2010) 054508,
  [\href{http://arxiv.org/abs/0912.3360}{{\tt arXiv:0912.3360}}].

\bibitem{Aarts:2011ax}
G.~Aarts, F.~A. James, E.~Seiler, and I.-O. Stamatescu, {\it {Complex Langevin:
  Etiology and Diagnostics of its Main Problem}},  {\em Eur. Phys. J.} {\bf
  C71} (2011) 1756, [\href{http://arxiv.org/abs/1101.3270}{{\tt
  arXiv:1101.3270}}].

\bibitem{Nishimura:2015pba}
J.~Nishimura and S.~Shimasaki, {\it {New insights into the problem with a
  singular drift term in the complex Langevin method}},  {\em Phys. Rev.} {\bf
  D92} (2015), no.~1 011501, [\href{http://arxiv.org/abs/1504.08359}{{\tt
  arXiv:1504.08359}}].

\bibitem{Nagata:2015uga}
K.~Nagata, J.~Nishimura, and S.~Shimasaki, {\it {Justification of the complex
  Langevin method with the gauge cooling procedure}},  {\em PTEP} {\bf 2016}
  (2016), no.~1 013B01, [\href{http://arxiv.org/abs/1508.02377}{{\tt
  arXiv:1508.02377}}].

\bibitem{Nagata:2016vkn}
K.~Nagata, J.~Nishimura, and S.~Shimasaki, {\it {Argument for justification of
  the complex Langevin method and the condition for correct convergence}},
  {\em Phys. Rev.} {\bf D94} (2016), no.~11 114515,
  [\href{http://arxiv.org/abs/1606.07627}{{\tt arXiv:1606.07627}}].

\bibitem{Ito:2016efb}
Y.~Ito and J.~Nishimura, {\it {The complex Langevin analysis of spontaneous
  symmetry breaking induced by complex fermion determinant}},  {\em JHEP} {\bf
  12} (2016) 009, [\href{http://arxiv.org/abs/1609.04501}{{\tt
  arXiv:1609.04501}}].

\bibitem{Anagnostopoulos:2017gos}
K.~N. Anagnostopoulos, T.~Azuma, Y.~Ito, J.~Nishimura, and S.~K. Papadoudis,
  {\it {Complex Langevin analysis of the spontaneous symmetry breaking in
  dimensionally reduced super Yang-Mills models}},  {\em JHEP} {\bf 02} (2018)
  151, [\href{http://arxiv.org/abs/1712.07562}{{\tt arXiv:1712.07562}}].

\bibitem{Aoyama:2010ry}
T.~Aoyama, J.~Nishimura, and T.~Okubo, {\it {Spontaneous breaking of the
  rotational symmetry in dimensionally reduced super Yang-Mills models}},  {\em
  Prog. Theor. Phys.} {\bf 125} (2011) 537--563,
  [\href{http://arxiv.org/abs/1007.0883}{{\tt arXiv:1007.0883}}].

\bibitem{Ito:2013ywa}
Y.~Ito, S.-W. Kim, Y.~Koizuka, J.~Nishimura, and A.~Tsuchiya, {\it {A
  renormalization group method for studying the early universe in the
  Lorentzian IIB matrix model}},  {\em PTEP} {\bf 2014} (2014), no.~8 083B01,
  [\href{http://arxiv.org/abs/1312.5415}{{\tt arXiv:1312.5415}}].

\bibitem{Ito:2015mxa}
Y.~Ito, J.~Nishimura, and A.~Tsuchiya, {\it {Power-law expansion of the
  Universe from the bosonic Lorentzian type IIB matrix model}},  {\em JHEP}
  {\bf 11} (2015) 070, [\href{http://arxiv.org/abs/1506.04795}{{\tt
  arXiv:1506.04795}}].

\bibitem{Ito:2017rcr}
Y.~Ito, J.~Nishimura, and A.~Tsuchiya, {\it {Universality and the dynamical
  space-time dimensionality in the Lorentzian type IIB matrix model}},  {\em
  JHEP} {\bf 03} (2017) 143, [\href{http://arxiv.org/abs/1701.07783}{{\tt
  arXiv:1701.07783}}].

\bibitem{Azuma:2017dcb}
T.~Azuma, Y.~Ito, J.~Nishimura, and A.~Tsuchiya, {\it {A new method for probing
  the late-time dynamics in the Lorentzian type IIB matrix model}},  {\em PTEP}
  {\bf 2017} (2017), no.~8 083B03, [\href{http://arxiv.org/abs/1705.07812}{{\tt
  arXiv:1705.07812}}].

\bibitem{Yang:2015vna}
H.~S. Yang, {\it {Emergent Spacetime and Cosmic Inflation I \& II}},
  \href{http://arxiv.org/abs/1503.00712}{{\tt arXiv:1503.00712}}.

\bibitem{Kim:2018mfv}
K.~K. Kim, S.~Koh, and H.~S. Yang, {\it {Expanding Universe and Dynamical
  Compactification Using Yang-Mills Instantons}},  {\em JHEP} {\bf 12} (2018)
  085, [\href{http://arxiv.org/abs/1810.12291}{{\tt arXiv:1810.12291}}].

\bibitem{Tomita:2015let}
K.~Tomita, {\it {Fluctuations of the cosmic background radiation appearing in
  the 10-dimensional cosmological model}},  {\em PTEP} {\bf 2015} (2015),
  no.~12 123E01, [\href{http://arxiv.org/abs/1511.08583}{{\tt
  arXiv:1511.08583}}].

\bibitem{workinprog}
T.~Aoki, M.~Hirasawa, Y.~Ito, J.~Nishimura, and A.~Tsuchiya, {\it {On the
  structure of the emergent 3d expanding space in the Lorentzian type IIB
  matrix model}},  \href{http://arxiv.org/abs/1904.05914}{{\tt
  arXiv:1904.05914}}.

\bibitem{Krauth:1998xh}
W.~Krauth, H.~Nicolai, and M.~Staudacher, {\it {Monte Carlo approach to M
  theory}},  {\em Phys. Lett.} {\bf B431} (1998) 31--41,
  [\href{http://arxiv.org/abs/hep-th/9803117}{{\tt hep-th/9803117}}].

\bibitem{Austing:2001pk}
P.~Austing and J.~F. Wheater, {\it {Convergent Yang-Mills matrix theories}},
  {\em JHEP} {\bf 04} (2001) 019,
  [\href{http://arxiv.org/abs/hep-th/0103159}{{\tt hep-th/0103159}}].

\bibitem{Anagnostopoulos:2013xga}
K.~N. Anagnostopoulos, T.~Azuma, and J.~Nishimura, {\it {Monte Carlo studies of
  the spontaneous rotational symmetry breaking in dimensionally reduced super
  Yang-Mills models}},  {\em JHEP} {\bf 11} (2013) 009,
  [\href{http://arxiv.org/abs/1306.6135}{{\tt arXiv:1306.6135}}].

\bibitem{Kim:2011ts}
S.-W. Kim, J.~Nishimura, and A.~Tsuchiya, {\it {Expanding universe as a
  classical solution in the Lorentzian matrix model for nonperturbative
  superstring theory}},  {\em Phys. Rev.} {\bf D86} (2012) 027901,
  [\href{http://arxiv.org/abs/1110.4803}{{\tt arXiv:1110.4803}}].

\bibitem{Kim:2012mw}
S.-W. Kim, J.~Nishimura, and A.~Tsuchiya, {\it {Late time behaviors of the
  expanding universe in the IIB matrix model}},  {\em JHEP} {\bf 10} (2012)
  147, [\href{http://arxiv.org/abs/1208.0711}{{\tt arXiv:1208.0711}}].

\bibitem{Chaney:2015ktw}
A.~Chaney, L.~Lu, and A.~Stern, {\it {Matrix Model Approach to Cosmology}},
  {\em Phys. Rev.} {\bf D93} (2016), no.~6 064074,
  [\href{http://arxiv.org/abs/1511.06816}{{\tt arXiv:1511.06816}}].

\bibitem{Chaney:2015mfa}
A.~Chaney, L.~Lu, and A.~Stern, {\it {Lorentzian Fuzzy Spheres}},  {\em Phys.
  Rev.} {\bf D92} (2015), no.~6 064021,
  [\href{http://arxiv.org/abs/1506.03505}{{\tt arXiv:1506.03505}}].

\bibitem{Chaney:2016npa}
A.~Chaney and A.~Stern, {\it {Fuzzy $CP^2$ spacetimes}},  {\em Phys. Rev.} {\bf
  D95} (2017), no.~4 046001, [\href{http://arxiv.org/abs/1612.01964}{{\tt
  arXiv:1612.01964}}].

\bibitem{Stern:2018wud}
A.~Stern and C.~Xu, {\it {Signature change in matrix model solutions}},  {\em
  Phys. Rev.} {\bf D98} (2018), no.~8 086015,
  [\href{http://arxiv.org/abs/1808.07963}{{\tt arXiv:1808.07963}}].

\bibitem{Steinacker:2017vqw}
H.~C. Steinacker, {\it {Cosmological space-times with resolved Big Bang in
  Yang-Mills matrix models}},  {\em JHEP} {\bf 02} (2018) 033,
  [\href{http://arxiv.org/abs/1709.10480}{{\tt arXiv:1709.10480}}].

\bibitem{Steinacker:2017bhb}
H.~C. Steinacker, {\it {Quantized open FRW cosmology from Yang-Mills matrix
  models}},  {\em Phys. Lett.} {\bf B782} (2018) 176--180,
  [\href{http://arxiv.org/abs/1710.11495}{{\tt arXiv:1710.11495}}].

\bibitem{Chatzistavrakidis:2011gs}
A.~Chatzistavrakidis, H.~Steinacker, and G.~Zoupanos, {\it {Intersecting branes
  and a standard model realization in matrix models}},  {\em JHEP} {\bf 09}
  (2011) 115, [\href{http://arxiv.org/abs/1107.0265}{{\tt arXiv:1107.0265}}].

\bibitem{Nishimura:2013moa}
J.~Nishimura and A.~Tsuchiya, {\it {Realizing chiral fermions in the type IIB
  matrix model at finite N}},  {\em JHEP} {\bf 12} (2013) 002,
  [\href{http://arxiv.org/abs/1305.5547}{{\tt arXiv:1305.5547}}].

\bibitem{Steinacker:2014fja}
H.~C. Steinacker and J.~Zahn, {\it {An extended standard model and its Higgs
  geometry from the matrix model}},  {\em PTEP} {\bf 2014} (2014), no.~8
  083B03, [\href{http://arxiv.org/abs/1401.2020}{{\tt arXiv:1401.2020}}].

\bibitem{Aoki:2014cya}
H.~Aoki, J.~Nishimura, and A.~Tsuchiya, {\it {Realizing three generations of
  the Standard Model fermions in the type IIB matrix model}},  {\em JHEP} {\bf
  05} (2014) 131, [\href{http://arxiv.org/abs/1401.7848}{{\tt
  arXiv:1401.7848}}].

\bibitem{on-going}
K.~Hatakeyama, A.~Matsumoto, J.~Nishimura, A.~Tsuchiya, and A.~Yosprakob {\!\!,
  in preparation}.

\end{thebibliography}\endgroup


\end{document}